\documentclass[journal=jpclcd,
                manuscript=letter,
                ]{achemso}

\setkeys{acs}{maxauthors = 100}

\usepackage{amsmath}
\usepackage{amssymb}
\usepackage{url}
\usepackage{hyperref}

\usepackage{xcolor}
\newcommand{\hl}[1]{\textcolor{black}{#1}}
\newcommand{\hhl}[1]{\textcolor{black}{#1}}

\author{Xiran Yang}
\affiliation{Institute of Theoretical Chemistry, Faculty of Chemistry, University of Vienna, \\W\"ahringer Stra\ss{}e 17, 1090 Vienna, Austria}
\alsoaffiliation{Research Platform on Accelerating Photoreaction Discovery (ViRAPID), University of Vienna, Währinger Straße 17, 1090 Vienna, Austria}
\alsoaffiliation{Vienna Doctoral School in Chemistry, University of Vienna, Währinger Straße 42, 1090 Vienna, Austria.}
\author{Madlen Maria Reiner}
\affiliation{Faculty of Physics, University of Vienna, Kolingasse 14-16, 1090 Vienna, Austria}
\alsoaffiliation{Research Platform on Accelerating Photoreaction Discovery (ViRAPID), University of Vienna, Währinger Straße 17, 1090 Vienna, Austria}
\alsoaffiliation{Vienna Doctoral School in Physics, University of Vienna, Boltzmanngasse 5, 1090 Vienna, Austria}
\author{Brigitta Bachmair}
\affiliation{Institute of Theoretical Chemistry, Faculty of Chemistry, University of Vienna, \\W\"ahringer Stra\ss{}e 17, 1090 Vienna, Austria}
\alsoaffiliation{Research Platform on Accelerating Photoreaction Discovery (ViRAPID), University of Vienna, Währinger Straße 17, 1090 Vienna, Austria}
\alsoaffiliation{Vienna Doctoral School in Chemistry, University of Vienna, Währinger Straße 42, 1090 Vienna, Austria.}
\author{Leticia Gonz\'alez}
\email{leticia.gonzalez@univie.ac.at}
\affiliation{Institute of Theoretical Chemistry, Faculty of Chemistry, University of Vienna, \\W\"ahringer Stra\ss{}e 17, 1090 Vienna, Austria}
\alsoaffiliation{Research Platform on Accelerating Photoreaction Discovery (ViRAPID), University of Vienna, Währinger Straße 17, 1090 Vienna, Austria}
\author{Johannes C. B. Dietschreit}
\email{johannes.dietschreit@univie.ac.at}
\affiliation{Institute of Theoretical Chemistry, Faculty of Chemistry, University of Vienna, \\W\"ahringer Stra\ss{}e 17, 1090 Vienna, Austria}
\alsoaffiliation{Research Platform on Accelerating Photoreaction Discovery (ViRAPID), University of Vienna, Währinger Straße 17, 1090 Vienna, Austria}
\author{Christoph Dellago}
\affiliation{Faculty of Physics, University of Vienna, Kolingasse 14-16, 1090 Vienna, Austria}
\alsoaffiliation{Research Platform on Accelerating Photoreaction Discovery (ViRAPID), University of Vienna, Währinger Straße 17, 1090 Vienna, Austria}
\email{christoph.dellago@univie.ac.at}

\title[NATPS]
  {NATPS: 
  Nonadiabatic Transition Path Sampling Using the Time-Reversible Mapping Approach to Surface Hopping}

\abbreviations{NAMD, TPS}
\keywords{nonadiabatic molecular dynamics, transition path sampling, enhanced sampling, mapping approach to surface hopping, excited-state rare events}

\begin{document}

\makeatletter
\setlength\acs@tocentry@height{4.45cm}
\setlength\acs@tocentry@width{8.25cm}
\makeatother
\begin{tocentry}
\centering
\includegraphics[width=8cm]{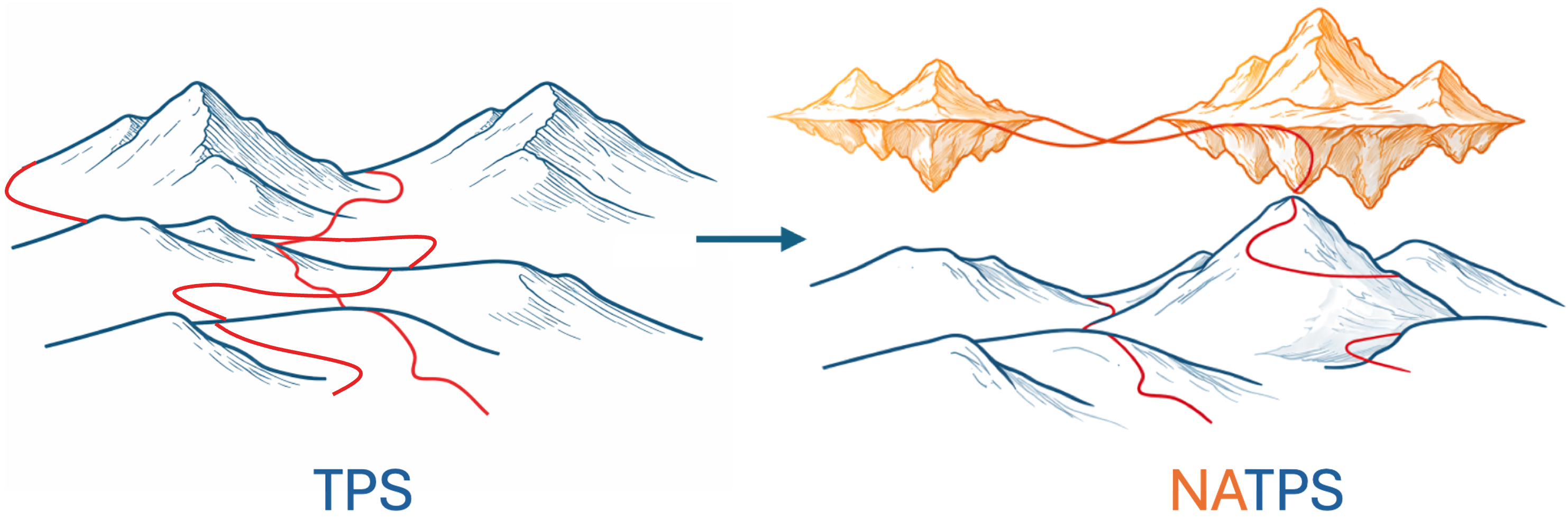}






\end{tocentry}

\begin{abstract}
  Rare nonadiabatic events play a central role in photochemistry but remain difficult to simulate because excited-state dynamics is computationally demanding and often stochastic. 
    Here we introduce a deterministic and time-reversible implementation of nonadiabatic dynamics that enables the application of transition path sampling (TPS) to excited-state processes. 
    Our approach builds on the mapping approach to surface hopping (MASH) and establishes the conditions required for path ensemble sampling, in particular, time reversibility and detailed balance. 
    The combination of MASH with the TPS framework gives rise to a new method termed nonadiabatic transition path sampling (NATPS).
    We demonstrate its capabilities on a model system of electronically coupled potential energy surfaces, where it efficiently generates ensembles of reactive trajectories and provides mechanistic insight into nonadiabatic pathways. 
    Compared with brute-force trajectory simulations and forward-flux sampling approaches, NATPS substantially reduces the computational effort required to obtain reactive trajectories.
\end{abstract}

\hl{Nonadiabatic processes are characterized by transitions between electronic states. 
They are central to modern photochemistry and molecular photophysics, encompassing a wide range of reactions like ring-opening, isomerization, and bond dissociation.~\cite{CrespoHernndez2004, Melchiorre2022}}
Although nonadiabatic transitions typically occur on femtosecond timescales,\cite{deNaldaBanares2014} the overall kinetics of excited-state processes may span many orders of magnitude.
\hl{The apparent conflict between the contrasting timescales can be attributed to the low transition probabilities from weak electronic couplings and the low accessibility of electronically coupled regions, which} may require thermally activated motion, diffusion through complex phase-space landscapes, or escape from long-lived excited-state minima.\cite{DomckeYarkonyKoppel2011, deNaldaBanares2014}

Consequently, many of the nonadiabatic processes manifest as kinetic rare events despite the ultrafast nature of the underlying microscopic dynamics. 
\hl{%
General examples include electron transfer\cite{Marcus1964_ET}, predissociation\cite{Kato1995, Sato2001},  photoisomerization\cite{Gida2002, Ma2024, Ma2025}, and intersystem crossing\cite{Penfold2018}, which can span a wide range of timescales, from ultrafast to slow regimes.
More concrete examples are the minor photodissociation channel of ammonia, NH$_3$ $\to$ NH + H$_2$ with a quantum yield of less than one percent, requiring tens of thousands of brute-force surface hopping trajectories to characterize statistically,\cite{Bachmair2025}
and the intersystem crossing in sulfur-substituted nucleobases, which involves long-lived triplet states whose formation and decay are difficult to access computationally due to the low probability of the relevant spin-forbidden nonadiabatic transitions.\cite{Duwal2025}
These examples highlight the need to redirect the computational effort on the rare reactive events themselves, rather than the long waiting times between them, regardless of how such waiting times arise.}

One way to achieve this judicious redirection is through enhanced sampling techniques, which can extract statistically meaningful information from a limited number of trajectory time steps.
In particular, trajectory-based enhanced sampling methods such as transition path sampling (TPS)\cite{Dellago1998, Bolhuis2000, Dellago2002,Dellago2009}, transition interface sampling\cite{vanErp2003, Moroni2004, vanErp2005}, and forward flux sampling (FFS)\cite{AllenWarrenTenWolde2005, AllenFrenkelTenWolde2006a, AllenFrenkelTenWolde2006b,Allen2009a} provide powerful tools to characterize rare events in the electronic ground state.
Path sampling techniques that generate both forward and backward trajectories usually require the underlying dynamics to be time reversible, to conserve phase-space volume, and to satisfy detailed balance in order to define a consistent and readily accessible trajectory probability measure. 
Unfortunately, most standard mixed quantum-classical approaches employed to describe nonadiabatic dynamics generally do not fulfill these conditions.\cite{Subotnik2015} In particular, fewest-switches surface hopping (FSSH)\cite{Tully1990} introduces stochastic switching rules for discrete electronic transitions that break microscopic reversibility and violate detailed balance\cite{Subotnik2015}.

Yet, some progress has recently been made toward rare-event sampling of nonadiabatic dynamics.
The earliest attempt to reconcile FSSH with TPS was introduced by Sherman and Corcelli,\cite{natpsgrandparent} who generated backward segments using a surrogate, history-independent hopping scheme and recovered the correct path probability by reweighting through forward retracing of the nuclear trajectory to reconstruct the true fewest-switches probabilities.
Some of us also  developed a nonadiabatic extension of forward flux sampling (NAFFS),\cite{Reiner2023} which is computationally efficient because FFS propagates trajectories forward, imposing less stringent constraints on the underlying dynamics.
In general, there is a pressing need for  nonadiabatic dynamics approaches that preserve the statistical-mechanical structure required for path sampling.

\begin{figure}
    \centering
    \includegraphics[width=0.8\linewidth]{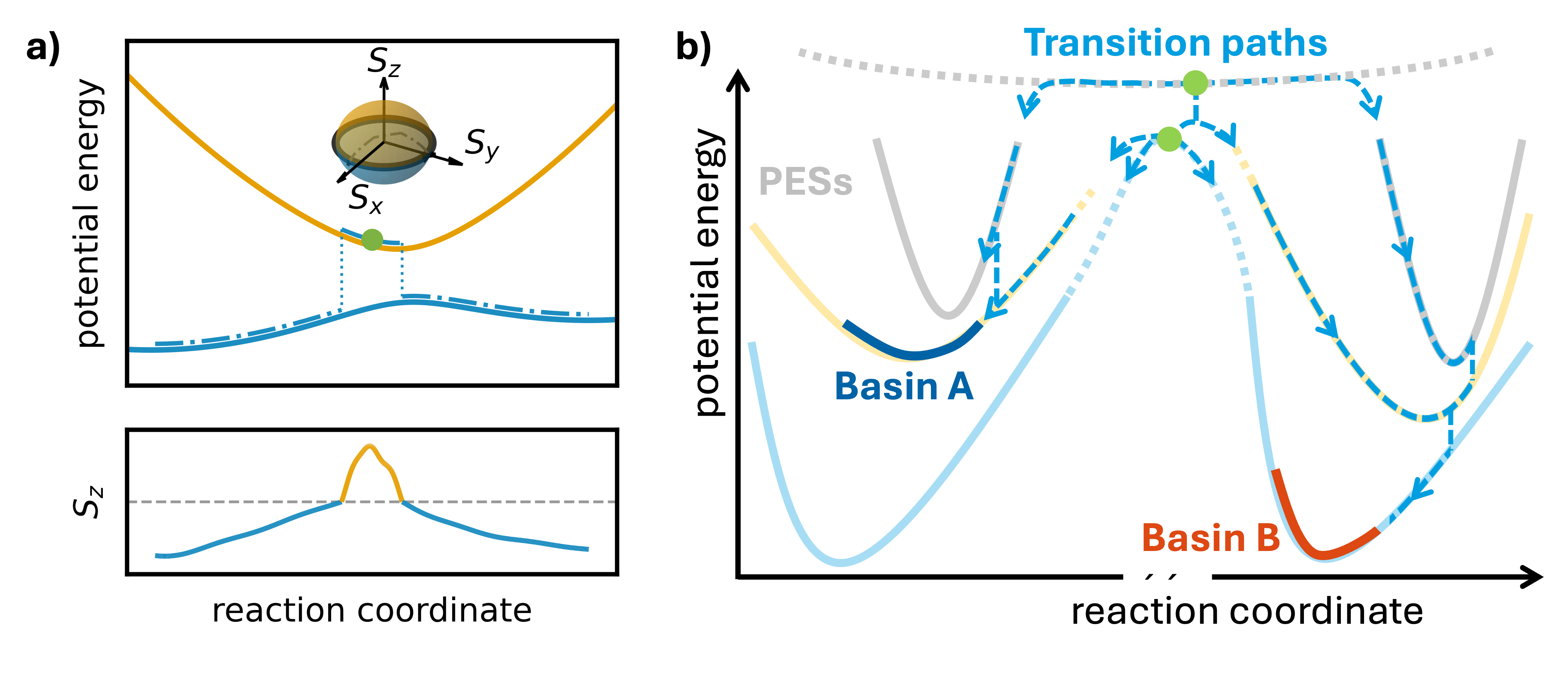}
    \caption{%
        a) Illustration of the spin-Bloch sphere used in the mapping approach to surface hopping (MASH) and a trajectory in a two-level system, initialized from the green circle. 
        The state is determined by the sign of the $z$-component of the spin-vector $\vec{S}$ depicted below.
        b) Illustration of nonadiabatic potential energy surfaces (PESs) with basin $\mathrm{A}$ marked in dark blue and basin $\mathrm{B}$ marked in red.
        Nonadiabatic transition path sampling (NATPS) attempts to generate paths that connect the two regions, by starting new trajectories from the shooting points (green dots) propagating the system forward and backward in time until both ends reach a different basin to form a reactive path.
        If no such path is found, then the shooting is re-attempted at a new shooting point from the most recent reactive path.
        Each trajectory propagated with MASH can nonadiabatically switch between the different PESs.
        } 
    \label{fig:NATPS_Illustration}
\end{figure}

In this Letter, we focus on the mapping approach to surface hopping (MASH)\cite{Mannouch2023} method to develop a TPS framework for nonadiabatic dynamics.
MASH provides a framework for nonadiabatic dynamics that fulfills detailed balance\cite{Amati2023} and time reversibility\cite{Geuther2025}. 
In MASH, the coefficients of the electronic wave function of a two-state system 
\begin{equation}
\left|{\Psi(t)}\right> = c_- \left|{\psi_-}\right> + c_+ \left|{\psi_+}\right>\ ,
\label{eq:elec_wavefunc_main}
\end{equation}
are mapped to a continuous spin vector on a Bloch sphere\cite{Lawrence2024} (see Fig.~\ref{fig:NATPS_Illustration}a)
\begin{equation}
S_x = 2\,\mathrm{Re}(c_{+}^{*} c_{-}) \ , \quad \mathrm{and} \quad
S_y = 2\,\mathrm{Im}(c_{+}^{*} c_{-}) \ , \quad \mathrm{and} \quad
S_z = \lvert c_{+} \rvert^2 - \lvert c_{-} \rvert^2 \ ,
\label{eq:spin_vector_from_coefficients}
\end{equation}
which is coupled to the classical nuclei.
In this way, MASH yields smooth, deterministic equations of motion for the combined nuclear–electronic phase space.

MASH is based on the assumption that two coupled electronic states can be represented by the energy function
\begin{equation}
E(\mathbf{q}, \mathbf{p}, \vec{S}) 
= \sum_{i=1}^N \frac{p_i^2}{2m_i} 
  + \bar{V}(\mathbf{q}) 
  + V_z(\mathbf{q}) \, \mathrm{sign}(S_z)\ ,
\label{eq:Epot_twostate_main}
\end{equation}
where $\mathbf{q} = (q_1, \dots, q_{N})^\top$ are the generalized nuclear coordinates and 
$\mathbf{p} = (p_1, \dots, p_{N})^\top$ the conjugate nuclear momenta with masses $m_i$ for a system with $N$ degrees of freedom. 
$\bar{V}(\mathbf{q})$ denotes the average, state-independent potential, and 
$V_z(\mathbf{q})$ represents the half difference between the two electronic potential energy surfaces (PESs).
$S_z$ is the $z$-component of the auxiliary spin vector $\vec{S} = (S_x, S_y, S_z)^\top$.\cite{Richardson2025,Geuther2025}
As a comparison, in FSSH, the energy function of the system is given by\cite{Mannouch2023}
\begin{equation}
    E(\mathbf{q}, \mathbf{p}, n_{\text{active}}) = \sum_{i=1}^N \frac{p_i^2}{2m_i} + \bar{V}(q) + V_z n_{\text{active}} \ ,
\end{equation}
with $n_{\text{active}} = \pm 1$ indicating the active state. 
The hopping probability within a time step $\Delta t$ is calculated from electronic state coefficients\cite{lawrence2023recoveringmarcustheoryrates}:
\begin{equation}
    \begin{cases}
        P_{-\to +} = \dfrac{\Delta t}{|c_-(t)|^2}\dfrac{\partial |c_+(t)|^2}{\partial t} \\
        P_{+\to -} = \dfrac{\Delta t}{|c_+(t)|^2}\dfrac{\partial |c_-(t)|^2}{\partial t}\ .
    \end{cases}
\end{equation}
Whereas in MASH the active state of the system is specified by the sign of the continuous $S_z$, FSSH has to rely on a probability measure and stochastic hopping.
A more detailed background on MASH is given in Section~S1 of the Supporting Information (SI) including the equations of motion for positions and momenta (Section~S1.1) as well as the auxiliary spin-vector (Section~S1.2).

The canonical Boltzmann distribution associated with Eq.~(\ref{eq:Epot_twostate_main}) is given by
\begin{equation}
\rho(\mathbf{q}, \mathbf{p}, \vec{S})
\propto
\exp\!\left[
 -\beta\!\left(
   \sum_i \frac{p_i^2}{2m_i}
   + \bar{V}(\mathbf{q})
   + V_z(\mathbf{q}) \, \mathrm{sign}(S_z)
 \right)
 \right]
\, \delta(|\vec{S}| - 1)\ .
\label{eq:pdf_main}
\end{equation}
Using the Fokker-Planck equation (Section~S1.4), it can be shown that the Boltzmann distribution, Eq.~(\ref{eq:pdf_main}), is stationary under the MASH dynamics. 
Here $\beta = 1/(k_\mathrm{B}T)$, where $k_\mathrm{B}$ denotes the Boltzmann constant and $T$ the absolute temperature.
The projection of the full Boltzmann distribution onto the $z$-axis of the spin vector gives the marginal probability density along $S_z$, which is uniform for each hemisphere (Section~S1.3)\cite{Richardson2025} and from which one can obtain by direct integration the free energy difference between the upper and lower electronic states\cite{dietschreit_how_2022}. 

For the application of MASH with TPS \cite{Dellago1998,Bolhuis2002}, the underlying dynamics must additionally satisfy time reversibility. 
We demonstrate (Section~S2) that using the MASH equations of motion both the nuclear degrees of freedom and the extended spin vector conserve phase-space volume.

Under time reversal, the phase-space variables transform according to their parity. 
For the nuclear degrees of freedom, this corresponds to the standard classical transformation: positions remain unchanged, while momenta inverse their sign. 
The auxiliary spin-vector exhibits a different symmetry: the $x$- and $z$-components are even under time reversal, whereas the $y$-component is odd (see Section~S3 for a detailed derivation).
With these parity relations, the MASH equations of motion are invariant under time reversal, \textit{i.e.}, the dynamics is microscopically reversible. 
This implies that for every trajectory in phase space, the time-reversed trajectory occurs with the same probability. 
Together with the fact that the Boltzmann distribution in Eq.~(\ref{eq:pdf_main}) is stationary under the dynamics, this establishes detailed balance for MASH (see Section~S4).

Although it has been noted previously that MASH obeys detailed balance\cite{Amati2023} and samples the Boltzmann distribution associated with Eq.~(\ref{eq:Epot_twostate_main})\cite{Mannouch2023}, here we provide a complete and rigorous derivation in Sections~S1 to~S4. 
In addition, we show that the propagation of the coefficient vector $\mathbf{c}$ and the spin-vector $\vec{S}$ are equivalent (Sections~S5 and~S6). 
Applying the time-reversal transformation to the spin-vector components in Eq.~(\ref{eq:spin_vector_from_coefficients}) yields the corresponding symmetry for the electronic coefficients (Section~S7),
\begin{equation}
c_i \rightarrow c_i^{*} .
\label{eq:inversion_coefficients}
\end{equation}
With respect to the electronic coefficients, NATPS does not introduce any additional approximations beyond those inherent to the underlying MASH dynamics. 
In principle, any correction scheme developed for MASH, such as decoherence corrections via quantum jumps, can be incorporated into NATPS. 
However, given the established accuracy of bare MASH for the system considered here~\cite{Richardson2025}, such extensions are not required.

In this work, we propagate the nuclear degrees of freedom using the velocity-Verlet algorithm, and we employ the local diabatization\cite{felixld} scheme to propagate the electronic coefficients $\mathbf{c}$ (details are given in Section~S8), where after each time propagation they are mapped onto the components of the spin vector $\vec{S}$ (see Eq.~(\ref{eq:spin_vector_from_coefficients})). 
In the absence of surface hops, the velocity-Verlet integrator guarantees time-reversible nuclear dynamics. 
However, in conventional surface-hopping implementations, a change of the electronic state is only detected after the completion of a nuclear time step. 
As a result, the potential energy surface governing the nuclear propagation during a hopping event depends on the direction of time (forward vs.\ backward). 
This asymmetry breaks time-reversal symmetry, even though the underlying MASH equations of motion are microscopically reversible.
To resolve this asymmetry, we adopt the piecewise continuous approach developed by Geuther and Richardson \cite{Geuther2025}. 
In this approach, potential hopping events are detected before the nuclear propagation step is completed. 
An iterative root search is used to determine the precise time at which the hop occurs. 
The system is then propagated with a reduced time step up to the hopping event, where the velocity rescaling or reflection is performed, followed by the propagation for the remainder of the original time step. 
This approach ensures that the nuclei traverse the correct potential energy surfaces regardless of time direction, thus restoring full time-reversibility.
This method assumes that at most one hop occurs per time step. 
That this variable time step implementation is crucial for time-reversibility is shown through numerical experiments in Section~S9.


TPS provides a general framework for sampling rare reactive trajectories without prior knowledge of the reaction coordinate.\cite{Dellago1998, Bolhuis2000, Dellago2002,Dellago2009,Escobedo2009}
The method is based on a statistical description of trajectories, which are then sampled using a Markov Chain Monte Carlo procedure. For a trajectory $X$ of length $L$, the path probability distribution $P$ is given by\cite{Haeupl2026}
\begin{equation}
P[X(L)] := \rho(\mathbf{\Gamma}_0)
\prod_{i=0}^{L-1} P(\mathbf{\Gamma}_i \to \mathbf{\Gamma}_{i+1}),
\label{eq:path_probability}
\end{equation}
where $\mathbf{\Gamma} = (\mathbf{q}^\top, \mathbf{p}^\top, S_x, S_y, S_z)^\top$ is a point in the extended phase space, $\rho(\mathbf{\Gamma}_0)$ is the equilibrium probability of the initial phase-space point $\mathbf{\Gamma}_0$, and $ P(\mathbf{\Gamma}_i \to \mathbf{\Gamma}_{i+1})$ is the short-time transition probability determined by the underlying dynamical propagator.
The conditional path probability for trajectories that start in basin $\mathrm{A}$ and end in basin $\mathrm{B}$ is then given by
\begin{equation}
P_{\text{AB}}[X(L)] = Z_{\text{AB}}^{-1} H_{\text{AB}}[X(L)] P[X(L)] \ ,
\label{eq:path_prob}
\end{equation}
where $H_{\text{AB}}[X(L)]$ is unity if and only if i) the first configuration of the trajectory $X(L)$ lies in basin A; ii) the last lies in basin B; iii) no other configurations are in either of the basins.\cite{Haeupl2026}
$Z_{\text{AB}}$ is the normalization factor.
\hhl{The path probability defined in Eq.~(\ref{eq:path_probability}) corresponds to an equilibrium ensemble in path space. 
Since the underlying MASH dynamics obeys detailed balance and samples the Boltzmann distribution (Eq.~(\ref{eq:pdf_main})), a sufficiently long brute-force trajectory generates an ensemble of trajectory segments distributed according to $P[X(L)]$. 
Within this ensemble, transition paths are those trajectories that satisfy the constraint $H_{\text{AB}}[X(L)] = 1$, \textit{i.e.}, paths connecting basin A to basin B. Although such transition paths necessarily traverse high-energy and low-probability regions of configuration space and therefore exhibit non-equilibrium behavior at the level of individual trajectories, their statistical weight is defined with respect to the underlying equilibrium path ensemble. 
Equilibrium in this context thus refers to the probability distribution over paths, rather than to individual configurations along a trajectory.}
Trajectories in this restricted path ensemble are sampled using a Monte Carlo (MC) random walk in path space.\cite{Buijsman2020}
Given an existing path $X^{(o)}$, a new path $X^{(n)}$ is proposed and accepted according to the Metropolis-Hastings criterion.
If (i) the dynamical propagator satisfies microscopic reversibility and (ii) the path-generation procedure obeys detailed balance, the acceptance probability is
\begin{equation}
\label{Pacc}
P_{\text{acc}}(X^{(o)} \to X^{(n)})
=
h_\text{A}(\mathbf{\Gamma}^{(n)}_{0})\,
h_\text{B}(\mathbf{\Gamma}^{(n)}_{L_n})
\,
\text{min}\left\{1, \frac{L_o}{L_n}\right\},
\end{equation}
where \hl{$L_o$ and $L_n$ are the lengths of the old and new path respectively, and} $h_\zeta(\mathbf{\Gamma})$ ($\zeta=\text{A, B}$) is the indicator function that gives 1 for any configuration belonging to basin $\zeta$ and otherwise 0. 
$\mathbf{\Gamma}^{(n)}_{0}$ and $\mathbf{\Gamma}^{(n)}_{L_n}$ denote the first and last configurations of the proposed trajectory.
\hl{It is important to note that $\mathrm{A}$ and $\mathrm{B}$ can be defined as regions of any extended phase space coordinate, or combinations thereof, including conditions on one or multiple electronic states.
This allows NATPS to be directly applied to processes where the reactant and product reside on different electronic states without any modification to the method.}

New trajectories are generated using a shooting algorithm.
At each MC step, a configuration is randomly selected from the current trajectory and perturbed.
The perturbation typically involves modifying the velocity vector associated with the selected configuration.
The specific velocity perturbation determines the statistical ensemble sampled by the path ensemble. For instance, sampling the microcanonical (NVE) ensemble requires that the kinetic energy of the selected configuration remains unchanged.
In one-dimensional systems, this restriction limits the velocity modification to a mere sign change, which significantly reduces the diversity of sampled trajectories.
In contrast, sampling the canonical (NVT) ensemble allows the velocity to fluctuate according to the Maxwell-Boltzmann distribution at a given temperature. A schematic representation of this procedure in combination with the MASH dynamics is shown in Figure \ref{fig:NATPS_Illustration}.

To ensure detailed balance of the path generation, new velocities $v'$ are obtained using the Uhlenbeck-Ornstein \cite{Uhlenbeck1930} perturbation scheme,
\begin{equation}
v' = \alpha v + \sqrt{1-\alpha^2}\,\Delta v \ ,
\label{eq:Ornstein-Uhlenbeck}
\end{equation}
where $v$ is the current velocity, $\Delta v$ is a random variable drawn from the Maxwell-Boltzmann velocity distribution at temperature $T$, and $\alpha \in (0, 1)$ is a predefined mixing parameter, here $\alpha=0.9$, as it smoothly perturbs the velocity but still allows for rapid equilibration of the path ensemble (see discussion of Fig.~\ref{fig:potential} below). 
\hhl{A proof that this velocity perturbation scheme satisfies detailed balance is given in Section~S10.
As a result, the stationary distribution of the Markov chain coincides with the equilibrium path ensemble restricted to reactive trajectories. 
This means that TPS does not alter the probability distribution of transition paths compared to brute-force sampling, but provides a more efficient means to sample them.}

After selecting and perturbing a configuration $\mathbf{\Gamma}_\tau$, the system is propagated both forward and backward in time from this configuration to generate a new trajectory.
The resulting path is accepted or rejected according to Eq.~(\ref{Pacc}).
If the proposed path is rejected, the previous trajectory is retained in the ensemble to maintain the correct path probability distribution.
This shooting procedure is repeated until the desired number of trajectories has been collected.


\begin{figure}[!h]
    \centering
    \includegraphics[width=\linewidth]{figures/Fig2.png}
    \caption{%
    a) Analytic potentials along the nuclear coordinate $q$, with lower state in blue and upper state in orange. Basins $\mathrm{A}$ and $\mathrm{B}$ are shaded in gray.
    b) \hl{An example of a} reactive trajectory connecting basins $\mathrm{A}$ and $\mathrm{B}$, with the color indicating the adiabatic PES the system evolves on.
    c) Corresponding time evolution of the spin-vector component $S_z$ \hl{for the exemplary trajectory shown in panel b}.
    Orange segments indicate portions of the trajectory evolving on the upper state.
    d) Lineage history of trajectories during ten successive Monte Carlo steps in path space.
    Earlier trajectories are shown as faint thin lines, while the most recent trajectories are plotted as solid thick lines.
    Shooting points, from which new trajectories are sampled, are highlighted as  green dots. 
    A step-by-step version of this figure can be found in the SI, Figure~S2.
    e) Autocorrelation function associated with the transition time in MC path space. On average, a statistically independent path emerges after every 10 random walks. This is visually hinted from d), judging from the similarities among the trajectories.
    }
    \label{fig:potential}
\end{figure}

To illustrate the applicability of NATPS, we consider the one-dimensional dynamics of a particle with the mass of a hydrogen atom evolving on a two-state potential formed by two coupled harmonic oscillators, as illustrated in Fig.~\ref{fig:potential}a. 
Despite its simplicity, this model captures the essential physics of two electronically coupled diabatic states \cite{Marcus1993, Marcus1964_ET}.
The diabatic Hamiltonian of the system is defined as
\begin{equation}
    \mathbf{H}(q) = \begin{bmatrix}
        \dfrac{\epsilon}{x_0^2}(q-x_0)^2 & V_c\\
        V_c & \dfrac{\epsilon}{x_0^2}(q+x_0)^2
    \end{bmatrix} \ ,
    \label{eq:1d_potential}
\end{equation}
where $\epsilon = 0.05$~Ha and $x_0 = 1.0~a_0$, all quantities expressed in atomic units.
The mass of the particle is fixed at 1836.15~$m_e$.
For convenience, the electronic coupling $V_c$ is reported in units of $\epsilon$.
It should be noted that these units are intended as relative measures within the analytical framework and do not imply a direct mapping onto specific molecular species.
The Hamiltonian and choice of parameters are identical to those reported in the publication on NAFFS \cite{Reiner2023}, except that the system here is strictly one-dimensional.
A discussion of further details of the analytical model can be found in Section~S11.

The off-diagonal coupling $V_c$ induces an avoided crossing between the adiabatic potential energy surfaces. 
Unless specified otherwise, the off-diagonal coupling is always $V_c = 0.2\,\epsilon = 0.01$~Ha, as defined in Eq.~(\ref{eq:1d_potential}).
Evaluating the energy gap at $q=0$ yields the minimum adiabatic splitting $g=2\,V_c$, which is directly related to the barrier height on the adiabatic lower-state surface,
\begin{equation}
    E_a = \left(1-\frac{V_c}{\epsilon}\right)\epsilon \ .
\end{equation}
This barrier $E_a$, which is equivalent to the thermal energy $k_{\mathrm{B}}T$ at 12,632~K, controls the rareness of the transition that we investigate in this model system. 
We define the left basin $\mathrm{A}$ to include any configuration in the adiabatic lower state with $q<-0.8$ and $\mathrm{B}$ to encompass lower state configurations with $q>0.8$, indicated by the shaded regions in Fig.~\ref{fig:potential}a.

To ensure the root search for the exact hopping moment remains robust, we utilize a time step of $\Delta t=5$~a.t.u.\,.
Furthermore, to make sure that we sample from the correct TPS ensemble, we always discard the first 1500 generated transition paths (15\% of the paths) as equilibration-phase (for a discussion, see Section~S12).

Figures~\ref{fig:potential}b and~c show the time evolution of an unbiased trajectory in coordinate space (subfigure b) and the $z$-value of the spin-vector (subfigure c), where the colors indicate the electronic state of the system over time. 
The trajectory originates in basin $\mathrm{A}$, changes to the upper state, then returns to basin $\mathrm{A}$, before transitioning to basin $\mathrm{B}$ via a longer visit to the upper state. 
Only the segment from 900 to 1400 time units would be considered to be a transition path, as it successfully connects the two basins.


To demonstrate that our implementation of MASH combined with TPS correctly samples nonadiabatic trajectories, we generate a total of 10,000 transition paths at a temperature of 12,000~K using the Uhlenbeck-Ornstein velocity perturbation scheme described in Eq.~(\ref{eq:Ornstein-Uhlenbeck}). 
\hl{The initial conditions for the trajectory from which the TPS sampling is started, were chosen as $v_{t=0} = -0.0087~a_0/\mathrm{a.t.u}$ (corresponding to a kinetic temperature of 44,000~K),  $q_{t=0} = 1.0~a_0$, and $\vec{S}$ pointing to the south pole. 
Any alternative set of initial conditions that generated a trajectory connecting regions A and B would also have been acceptable.
Shooting points are selected uniformly at random along a previously identified reactive trajectory. 
Then, only the velocities are perturbed, while the positions and electronic coefficients are retained from the preceding trajectory.}

Analysis of the transition paths shows that $35\%$ of the ensemble at 12,000~K display at least two nonadiabatic transitions (upward and downward surface hops), confirming that the chosen \hl{parameters $T$ and $V_c$} are suitable for sampling both adiabatic and nonadiabatic transition paths.

The evolution of the ensemble in path space is illustrated by the path lineage history shown in Fig.~\ref{fig:potential}d. 
Earlier trajectories are displayed as faint, thin lines, while more recently accepted trajectories are rendered as thicker solid lines.
Consecutive paths in the ensemble exhibit significant correlation, as each new trajectory is generated from a modification of a previous one. 
Such correlations are expected for path sampling algorithms and imply that consecutive paths cannot be treated as statistically independent samples.
To quantify the correlation between sequential trajectories, we analyze their transition times, defined as the duration required for a trajectory to travel between basins $\mathrm{A}$ and $\mathrm{B}$. 
The degree of correlation can be measured using the transition-time autocorrelation function,

\begin{equation}
    C(n) = \frac{\sum_{t=0}^{N-n-1} (\tau(t) - \langle{\tau}\rangle)(\tau(t+n)-\langle{\tau}\rangle)}{(N-n)\sigma_\tau^2} \ ,
\end{equation}
as shown in Fig.~\ref{fig:potential}e. Here, $t$ and $n$ are steps in the MC path space, and $\sigma^2_\tau$ is the variance of the transition time.

For our settings, the decorrelation length is approximately 10~MC steps. 
Consequently, a TPS simulation that generates 10,000 trajectories yields an effective sample size of roughly 1000 statistically independent paths. 
This high ratio of independent paths to total MC steps highlights the efficiency of the NATPS sampling algorithm in exploring the nonadiabatic path space.

\begin{figure}[!h]
    \centering
    \includegraphics[width=\linewidth]{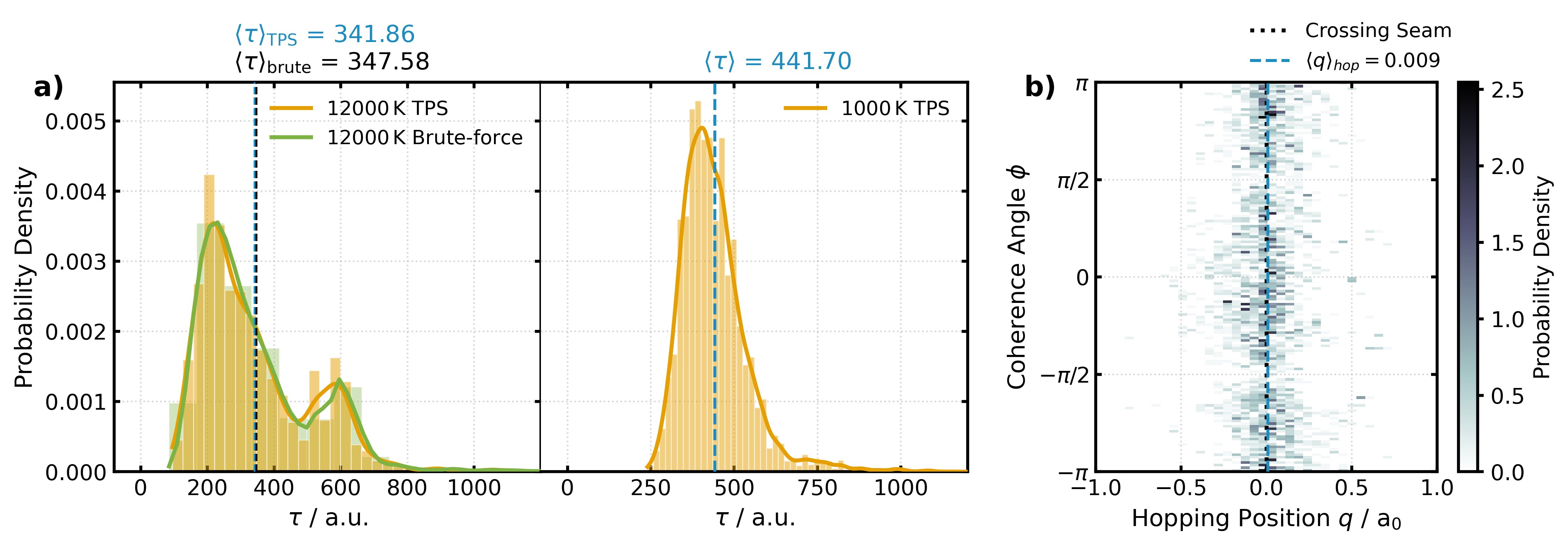}
    \caption{a) Probability distribution of transition times $\tau$ at 12,000~K (left) and 1000~K (right). 
    The 12,000~K transition path sampling (TPS) simulation result (orange) is overlaid with the brute-force simulation (green) with the initial conditions sampled from the stationary MASH distribution shown in Eq. (\ref{eq:pdf_main}) at the same temperature.
    Average transition times (blue for TPS and black for the brute-force simulation) are indicated above each graph. Kernel density envelopes are placed over the distributions to provide visual guidance on the general trend.
    b) Distribution of hopping positions \hl{and coherence angles} at 12,000~K. \hl{The coherence angle $\phi$ is defined as the arc tangent of the spin components $S_y$ and $S_x$.} Hops are concentrated symmetrically near the crossing seam $q=0$ \hl{ but have no dependence on $\phi$.} Any apparent asymmetries arise from finite sampling.
    Average hop position is indicated at the top.}
    \label{fig:ttdist12000}
\end{figure}

Beyond providing statistically independent trajectories, the TPS ensemble also contains mechanistic information about the reaction pathways. 
In particular, the distribution of transition times offers insight into the dynamical processes underlying the reactive trajectories. 
As shown in Fig.~\ref{fig:ttdist12000}a, the transition-time distribution at 12,000~K from the TPS simulation is right-skewed and bimodal.
This particular shape has two root causes.
First, there exists a minimum time imposed by the distance between the two basins and the kinetic energy of the system, but there is no strict upper bound except for the maximum path length used in the simulation.
Second, there are two mechanistically different pathways, adiabatic and nonadiabatic, present in the TPS ensemble. 
The nonadiabatic paths are typically slower, as they are trapped in the upper state for some time before connecting the two basins, which are strictly located in the lower state. 
In contrast, at the ``cooler'' temperature of 1000~K, the available kinetic energy is insufficient to induce upward hops to the excited state. 
Consequently, only adiabatic pathways remain, and the transition-time distribution becomes unimodal, as shown in the right panel of Fig.~\ref{fig:ttdist12000}a.

To validate our findings from the TPS simulation with the sampling of nonadiabatic transition paths, we carry out a brute-force simulation using the MASH dynamics by drawing initial conditions from the stationary MASH distribution shown in Eq. (\ref{eq:pdf_main}) at $T=12,000$~K. The result is overlaid with the one from TPS in the left panel of Fig.~\ref{fig:ttdist12000}a, showing near perfect agreement with each other.

Finally, TPS simulations also provide direct mechanistic insight into the location of nonadiabatic transitions. 
\hl{For the present one-dimensional model, this information is contained in the distribution of the nuclear geometries and electronic coefficients at the moment of hopping. Since at this time step $S_z = 0$, to express $S_x$ and $S_y$ compactly, we define
\begin{equation}
    \phi = \arctan2\left(S_y, S_x\right)
\end{equation}
as the angle of coherence. As shown in Fig.~\ref{fig:ttdist12000}b, the distribution is centered symmetrically around the crossing seam located at $q=0$, where the energy gap between the adiabatic states is minimal and the nonadiabatic coupling is the strongest. 
Since MASH only uses $S_z$ to determine a hop, the distribution of $\phi$ is uniform. 
We note that $\tan \phi$ is undefined at $\phi=\pm \frac{\pi}{2}$, creating artificial nodes at these positions in Fig.~\ref{fig:ttdist12000}b.}


Although TPS is not strictly required to obtain quantities such as transition-time distributions or hopping geometries, its primary advantage lies in its ability to efficiently sample rare reactive trajectories compared to brute-force molecular dynamics simulations. 
To quantify this advantage, we define the efficiency ratio as the average number of time steps spent to obtain one transition path.
Table~\ref{tab:ERratio} reports the efficiency ratio for NATPS and brute-force MASH simulations at different effective barrier heights $E_a$, as modulated by the temperature.

\begin{table}[tb]
\centering
\caption{Efficiency ratio $\mathrm{ER} = n(\text{total steps generated})/n(\text{transition paths})$.
The effective barrier height $E_a$ is expressed in units of $k_\mathrm{B}T$ and is controlled by varying the ensemble temperature.  
Larger values of $E_a$ therefore correspond to rarer reactive events.}
\label{tab:ERratio}

\begin{tabular}{c c c}
\hline
$E_a$ ($k_\mathrm{B}T$) & ER(NATPS) & ER(brute-force MASH)  \\
\hline
3  & 199 & 7752  \\
4  & 173 & 17,710  \\
5  & 195 & 72,115 \\
6  & 201 & 122,951 \\
10 & 186 & $>$15,000,000 (total simulated steps) \\
\hline
\end{tabular}
\end{table}

Across all investigated barrier heights, NATPS consistently requires significantly fewer integration steps to generate reactive trajectories than brute-force simulations.
For the largest barrier ($E_a = 10\,k_BT$), no transition paths are found using the brute-force MASH simulation.
While the efficiency of brute-force dynamics decreases rapidly as the barrier height increases, the computational cost of NATPS remains nearly constant across the tested range, indicating that the algorithm efficiently samples reactive trajectories without suffering from the exponential slowdown typical of brute-force simulations.
Note that NATPS as carried out here produces a transition path ensemble but lacks basin population or flux information. 
Waiting times spent in the basins are not obtained, so it is unknown how frequently the transition occurs in the unbiased system. 
As a result, some kinetic properties, such as the reaction rate constant, are not accessible.
The brute-force approach, albeit inefficient, can provide such information.

\begin{figure}
    \centering
    \includegraphics[width=0.5\linewidth]{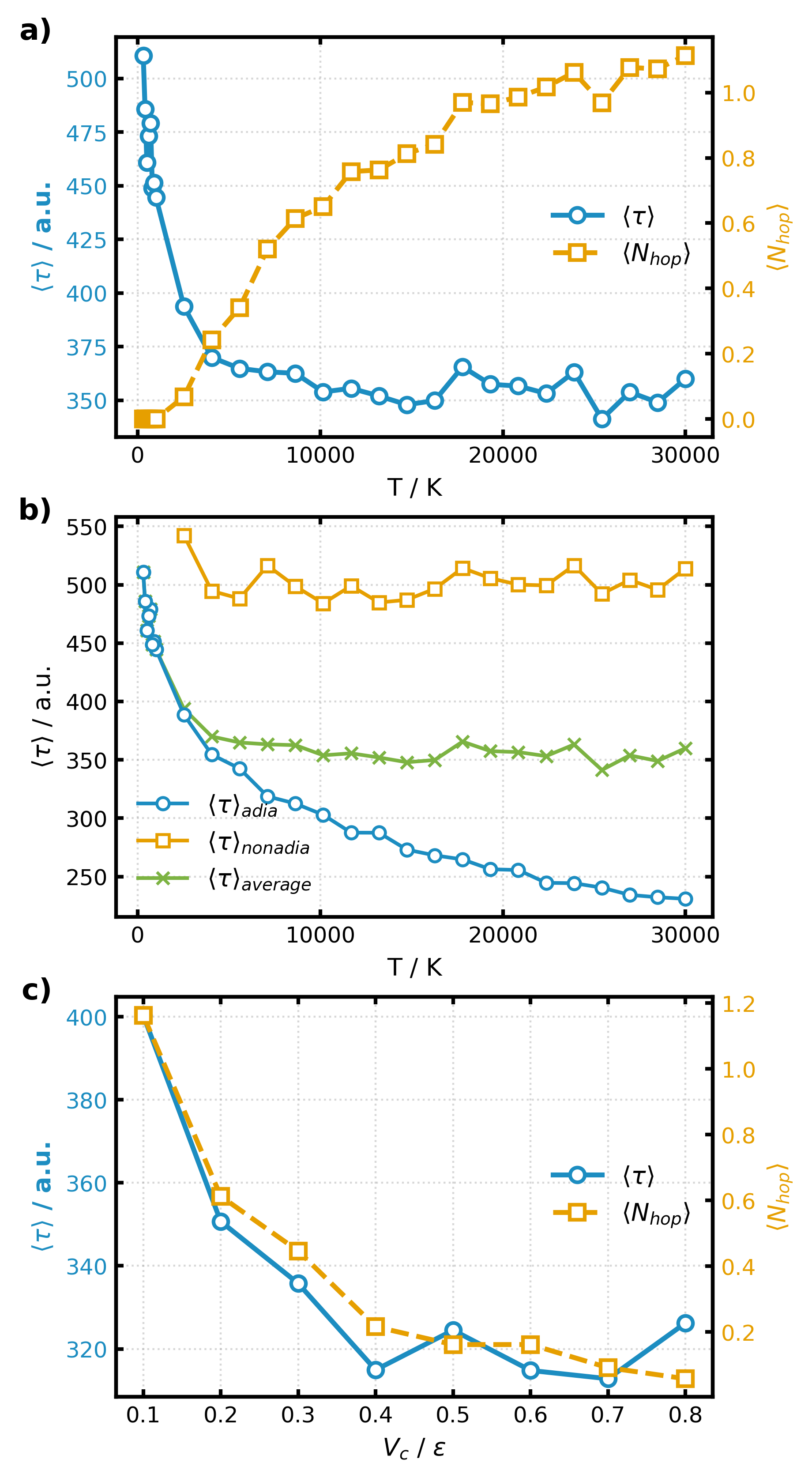}
    \caption{%
        Temperature dependence of TPS ensemble statistics showing a) mean transition time $\langle \tau \rangle$ and mean number of hops $\langle N_\mathrm{hop} \rangle$, and b) mean transition time for adiabatic and nonadiabatic paths separately,
        simulated at $V_c = 0.2\,\epsilon$. 
        \hl{The size of the ensemble $N_{\text{TPS}}$ increases with temperature such that the ratio between the standard deviation of transition times and $\sqrt{N_{\text{TPS}}}$ for a given temperature is constant.}
        c) Dependence of mean transition time $\langle \tau \rangle$ and mean number of hops $\langle N_\mathrm{hop} \rangle$ on the electronic coupling $V_c$, simulated at 10,000~K.
    }
    \label{fig:MTTMNTSTD_T}
\end{figure}

The bimodality observed in the right panel of Fig.~\ref{fig:ttdist12000}a suggests that the composition of the TPS ensemble depends strongly on the extent of nonadiabaticity of the system. 
We therefore examine how the ensemble properties change with the temperature $T$ and the electronic coupling $V_c$.
%
The coupling is fixed at $V_c = 0.2\,\epsilon$ and the temperature is varied between $300$ and $30,000$~K. 
To ensure stable equilibration across this wide temperature range, we employ a path-space annealing procedure: sampling begins at the highest temperature and the final trajectory of each ensemble is used to initialize the simulation at the next lower temperature.
The resulting temperature dependence of the mean transition time and the mean number of hops are shown in Fig.~\ref{fig:MTTMNTSTD_T}. \hl{We notice that as temperature increases, the distribution of transition times broadens and thus the standard deviation $\sigma_\tau (T)$ increases. To ensure reliable statistics from the path ensemble, we adapt the ensemble size $N_{\text{TPS}}$ so as to keep the ratio $\frac{\sigma_\tau (T)}{\sqrt{N_{\text{TPS}}}}$ approximately constant.}

In the low temperature range (up to 2000~K), the mean transition time decreases rapidly as $T$ increases, reflecting the larger kinetic energy available for barrier crossing. 
However, this trend weakens significantly at higher temperatures.
To understand this behavior, we separate trajectories into purely adiabatic paths and paths involving surface hops (Fig.~\ref{fig:MTTMNTSTD_T}b).
Adiabatic trajectories shorten steadily with $T$, as expected for increasingly ballistic motion. 
In contrast, nonadiabatic trajectories show only a weak temperature dependence. 
This behavior originates from excited state trapping: once a trajectory hops to the excited state, the time required to return to the ground state is primarily governed by the electronic coupling rather than the nuclear momentum. 
As a result, the excited-state lifetime becomes the dominant contribution to the total transition time, producing an effective kinetic plateau.

\begin{figure}
    \centering
    \includegraphics[width=0.8\linewidth]{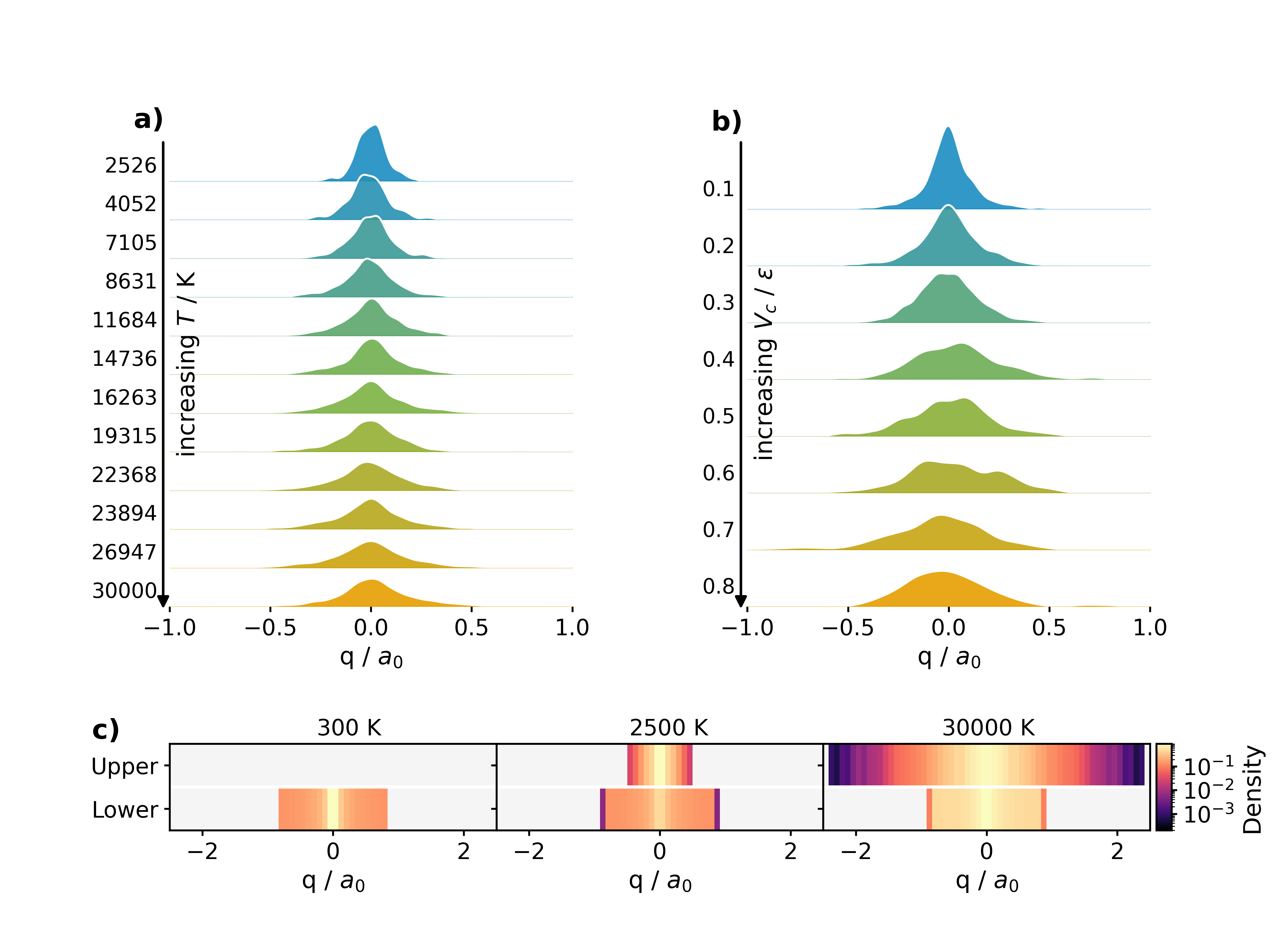}
    \caption{%
        Distribution of hopping positions for a) varying temperature and b) varying electronic coupling. c) State-resolved path density at 300~K, 2500~K, and 30,000~K.
        }
    \label{fig:3dhopdist_pathdensity}
\end{figure}

The influence of temperature is also visible in the spatial statistics of the TPS ensemble. 
Figure~\ref{fig:3dhopdist_pathdensity}a shows the distribution of hopping positions along the reaction coordinate.
At low temperatures, hops occur predominantly near the avoided crossing where the energy gap is smallest. 
As the temperature increases, larger nuclear momenta allow hops farther away from the crossing seam, leading to a gradual broadening of the hopping distribution.
This trend is also reflected in the state-resolved path densities at different temperatures shown in Fig.~\ref{fig:3dhopdist_pathdensity}c. 
Higher temperatures promote trajectories that explore larger regions of the higher-state surface before returning to the lower state.

We next examine the influence of the electronic coupling $V_c$, fixing the temperature at 10,000~K to ensure sufficient sampling of nonadiabatic paths.
Reducing $V_c$ lowers the adiabatic energy gap and increases the nonadiabatic coupling, thereby promoting surface hops. 
Conversely, increasing $V_c$ suppresses nonadiabatic transitions.
Figure~\ref{fig:MTTMNTSTD_T}c shows the resulting changes in the mean transition time and the mean number of hops as a function of the off-diagonal element $V_c$.

While increasing temperature and decreasing $V_c$ both promote hopping events, they do so through fundamentally different mechanisms. 
Temperature increases the available nuclear kinetic energy, whereas the coupling strength directly controls the electronic transition probability.
This distinction is reflected in the hopping distributions shown in Fig.~\ref{fig:3dhopdist_pathdensity}b. 
The temperature-induced broadening of the hopping distribution is gradual, whereas variations in $V_c$ lead to much stronger changes because the degree of spatial localization of the nonadiabatic coupling depends directly on the electronic Hamiltonian. 
Since the nonadiabatic coupling magnitude is sharply peaked near the crossing seam for small $V_c$ (for details see Section~S11), we obtain highly localized hopping regions, while larger couplings smear out the hopping distribution over a wider spatial range.


In conclusion, we introduced an NATPS framework based on MASH. 
By implementing a time-reversible propagation scheme within the MASH dynamics, we obtain a deterministic nonadiabatic dynamics method that satisfies detailed balance and is therefore compatible with trajectory-based path sampling algorithms. 
This is possible, as MASH allows the application of TPS methods that rely on backward propagation of partial trajectories, which are not generally accessible with stochastic surface-hopping dynamics.
\hhl{We note that while this manuscript was under revision, a complementary approach combining MASH and TPS has been published.\cite{ghamari2026nonadiabaticrareeventstransitionpath}}

\hl{In principle, any nonadiabatic dynamics method that supports time-reversible propagation and obeys detailed balance with respect to the equilibrium distribution can be combined with TPS to yield a nonadiabatic transition path sampling scheme.
We selected MASH specifically because it rigorously satisfies both of these  requirements.
By building on a particular dynamics method, however, we naturally inherit both its strengths and its limitations.
In its original formulation, MASH is strictly defined for two-level systems;  extensions to multiple electronic states are possible via approximations such as  unSMASH\cite{Lawrence2024} or multi-state MASH, also termed MISH,\cite{Runeson2023} and combining these with NATPS is a natural  direction for future work.
While the present work uses a one-dimensional model for clarity of presentation,  NATPS makes no assumptions about the number of nuclear degrees of freedom, and  the implementation can be applied directly to multidimensional systems.
} 

\hl{For the presented model system our NATPS implementation yields} path ensembles that reproduce the expected mechanistic features of the system, including the distribution of hopping geometries near the avoided crossing and the presence of distinct adiabatic and nonadiabatic reaction pathways. 
Analysis of the path statistics further shows that NATPS efficiently generates statistically independent reactive trajectories and provides direct access to mechanistic observables.
In addition, we find that variations in temperature primarily affect the nuclear kinetic energy and thereby the frequency of surface hops, while changes in the electronic coupling modify the spatial localization and probability of nonadiabatic transitions. 

\hl{Regarding the efficiency of NATPS we have to note that, like TPS in general, it performs best when the reactive trajectories are short compared to the waiting time between reactive events, as is the case for the studied model.
For diffusion-limited nonadiabatic processes, where the transition paths themselves are long, the efficiency advantage is reduced, though the method remains applicable.
However, for ultrafast processes, which are effectively barrierless, TPS does not provide any added benefit.}

Overall, NATPS offers a practical framework for studying rare nonadiabatic transitions using path sampling techniques. 
Because the method is based on deterministic dynamics that preserves the statistical-mechanical structure required for trajectory sampling, it provides a natural route toward applying advanced path ensemble methods to electronically excited-state processes in more complex molecular systems.
\hl{Future work will focus on the implementation in nonadiabatic dynamics packages such as SHARC\cite{SHARC4,Mai2018} for applying NATPS to multidimensional molecular systems, and combining it with techniques such as transition interface-based sampling (TIS)\cite{vanErp2003,vanErp2005} for the calculation of nonadiabatic rate constants.}

\begin{acknowledgement}
The authors thank Jeremy Richardson, Jonathan Mannouch, \hl{and Timothy Georges} for fruitful discussions. 
This work is funded by the University of Vienna within the framework of the research platform ViRAPID. 
Support of the Vienna Doctoral School in Physics (VDSP) is gratefully acknowledged. 

\end{acknowledgement}

\begin{suppinfo}

Derivations of the statistical-mechanical properties of the MASH dynamics, including the invariant distribution, phase-space volume conservation, time-reversal symmetry, and detailed balance. 
Relations between the spin-vector formulation and electronic wavefunction coefficients, including the treatment of frustrated hops and use of local diabatization. 
Additional theoretical details, numerical tests of time reversibility and integration accuracy, analytical properties of the model system (adiabatic energies and nonadiabatic couplings). 
The Supporting Information is available free of charge via the Internet at http://pubs.acs.org.

The code used to generate all data in this paper has been published at \url{https://github.com/Chikakoyanagida/NATPS}. 
\end{suppinfo}

\section*{Author Contributions}
\textbf{X.Y.}: Software -- lead, Investigation -- lead, Formal analysis -- lead, Visualization -- lead, Writing -- original draft.
\textbf{M.M.R.}: Investigation, Formal analysis, Software -- supporting, Writing -- original draft.
\textbf{B.B.}: Writing -- review \& editing.
\textbf{L.G.}: Conceptualization, Supervision, Funding acquisition -- equal, Writing -- review \& editing.
\textbf{J.C.B.D.}: Supervision, Software -- supporting, Validation, Visualization, Writing -- original draft.
\textbf{C.D.}: Conceptualization, Methodology, Supervision, Funding acquisition -- equal, Writing -- review \& editing.

\bibliography{references}

\end{document}